\documentclass[12pt]{article}
\usepackage[reqno]{amsmath}
\usepackage{epsfig}
\usepackage{array}
\usepackage{float}
\usepackage{lscape,graphicx}
\usepackage{graphics}
\usepackage{amssymb}
\usepackage{color}

\usepackage{array}
\newcolumntype{L}[1]{>{\raggedright\let\newline\\\arraybackslash\hspace{0pt}}p{#1}}
\newcolumntype{C}[1]{>{\centering\let\newline\\\arraybackslash\hspace{0pt}}p{#1}}
\newcolumntype{R}[1]{>{\raggedleft\let\newline\\\arraybackslash\hspace{0pt}}p{#1}}

\newcommand{\nonubb}{\mbox{$ 0\nu\beta\beta $ }}
\newcommand{\hs}{\hspace*{0.5cm}}

\newcommand{\be}{\begin{equation}}
\newcommand{\ee}{\end{equation}}
\newcommand{\bea}{\begin{eqnarray}}
\newcommand{\eea}{\end{eqnarray}}

\newcommand{\ea}{\end{array}}

\newcommand{\bad}{\begin{array}{ccc}}

\newcommand{\ba}{\begin{array}{c}}

\newcommand{\crn}{\nonumber \\}

\begin{document}

\bibliographystyle{ieeetr}

\begin{center}
\mathversion{bold}
{\bf{\large Neutrinoless double beta decay in a flipped trinification model with left-right symmetry}}
\mathversion{normal}

\vspace{0.4cm}
D. N. Dinh$\mbox{}^{a)}$\\
%
\vspace{0.1cm}
$\mbox{}^{a)}${\em Institute of Physics, Vietnam Academy
of Science and Technology, \\
10 Dao Tan, Ba Dinh, Hanoi, Vietnam.\\
}
%
\end{center}
\begin{abstract}
We investigate neutrinoless double beta ($0\nu\beta\beta$) decay within an extended Standard Model framework based on the 
$SU(3)_C\otimes SU(3)_L\otimes SU(3)_R \otimes U(1)_X$ gauge symmetry, commonly referred to as the filliped trinification model. 
Following a concise overview of the theoretical structure, we derive expressions for the effective Majorana mass 
and half-life associated with the dominant decay channels. A comprehensive numerical analysis is then carried out, 
accompanied by a discussion of the phenomenological implications.
\end{abstract}

\section{Introduction}
The Left-Right Symmetric Model (LRSM) \cite{Mohapatra:1974gc, Senjanovic:1975rk, Pati:1974yy, Senjanovic:1978ev} provides a compelling framework for physics beyond the Standard Model. This theoretical construction addresses two fundamental issues: the origin of maximal parity violation in weak interactions and the generation of small neutrino masses through the seesaw mechanism, facilitated by the inclusion of right-handed neutrino states. Furthermore, if left-right symmetry is broken at a scale of a few TeV, the model predicts a rich phenomenology of new physics. The nature of neutrino masses within LRSM depends on the symmetry-breaking mechanism, which may involve Higgs doublets, Higgs triplets, or combinations thereof. In scenarios with Higgs triplets, neutrino masses arise through a mixed seesaw mechanism (Type I and Type II), leading naturally to Majorana neutrinos. This framework also permits lepton flavor violation (LFV) and lepton number violation (LNV), with particular relevance to processes such as neutrinoless double beta decay ($0\nu\beta\beta$), which constitutes a central focus of this research.\\

Despite its advantages, the Left-Right Symmetric Model based on the gauge group 
$SU(2)_L\otimes SU(2)_L$ exhibits certain limitations. In particular, it does not provide an explanation for the number of fermion generations and achieves only limited success in addressing the nature of dark matter \cite{Bezrukov:2009th, Nemevsek:2012cd, Barry:2014ika, Heeck:2015qra, Garcia-Cely:2015quu, Cirelli:2005uq, Berlin:2016eem}. A promising alternative direction involves extending the electroweak sector from $SU(3)_C\otimes SU(2)_L\otimes U(1)$ to the larger gauge group $SU(3)_C\otimes SU(3)_L\otimes U(1)_X$. This class of models, initially proposed by Pisano and Frampton \cite{Singer:1980sw, Pisano:1992bxx, Frampton:1992wt} and subsequently developed by numerous authors \cite{Foot:1992rh, Montero:1992jk, Foot:1994ym, Dong:2006mg, Boucenna:2015zwa, Valle:2016kyz, Dong:2013wca, Dong:2014wsa, Huong:2015dwa, Huong:2016ybt}, has proven successful in addressing several outstanding problems in particle physics. Notably, the 
$SU(3)_C\otimes SU(3)_L\otimes U(1)_X$ framework offers natural explanations for the origin of neutrino masses, the existence and properties of dark matter, and the observed number of fermion generations.\\

In this research, we work on a theoretical framework that combines the features of Left-Right Symmetric Models and 3-3-1 models, constructed on the extended gauge group $SU(3)_C\otimes SU(3)_L\otimes SU(3)_R \otimes U(1)_X$. The model under consideration, known as flipped trinification \cite{Dong:2017zxo, Dinh:2019jdg}, inherits the attractive properties of both classes of models. Within this scenario, flavor-violating interactions naturally arise in both the quark and lepton sectors. Moreover, the characteristic scale of new physics is shown to be realizable at the order of ten TeV, thereby offering a rich phenomenology. Such a scale not only allows for observable low-energy effects but also provides distinctive signatures that may be probed at current collider experiments.\\

This study focuses on the lepton number violating process of neutrinoless double beta decay ($0\nu\beta\beta$) \cite{Schechter:1981bd}, in which two neutrons decay simultaneously into two protons and two electrons without the emission of neutrinos. Such a process can be mediated either by the exchange of light left-handed neutrinos with $W$-bosons, referred to as the standard (or fundamental) mechanism, or through the involvement of exotic particles, representing new physics contributions. Within the standard mechanism, current experimental sensitivity to 
$0\nu\beta\beta$ decay is limited to scenarios with quasi-degenerate light neutrino masses \cite{Bilenky:2001rz}, which have been excluded by cosmological observations \cite{Planck:2018vyg, RoyChoudhury:2018gay, RoyChoudhury:2019hls, Gerbino:2022nvz}. Consequently, any future experimental evidence of $0\nu\beta\beta$ decay would strongly indicate contributions from new physics. Current experimental bounds are: the GERDA collaboration reports a lower limit on the half-life of 
${\rm Ge^{76}}$, $T_{1/2}^{0\nu}>1.8\times 10^{26}$ years at $90\%$ C.L. \cite{GERDA:2020xhi}, corresponding to an upper bound on the effective Majorana mass of $m_{\beta\beta}^{0\nu}<(0.08-0.18)$ eV, subject to uncertainties in nuclear matrix elements. Similarly, EXO-200 \cite{EXO-200:2019rkq} and KamLAND-Zen \cite{KamLAND-Zen:2016pfg, KamLAND-Zen:2022tow}, both employing the isotope 
${\rm Xe^{136}}$, have established lower limits of $T_{1/2}^{0\nu}>3.5\times 10^{25}$ years and 
$T_{1/2}^{0\nu}>1.07\times 10^{26}$ years, respectively. These results, correspondingly, translate into constraints on the effective mass of 
$m_{\beta\beta}^{0\nu}<(0.093-0.286)$ eV and $m_{\beta\beta}^{0\nu}<(0.061-0.165)$ eV.\\

The present work examines the $0\nu\beta\beta$ decay and provides a detailed analysis of the contributions from new physics within the flipped trinification model \cite{Dong:2017zxo}. The paper is organized as follows. Section 2 presents a brief review of the model, including the Lagrangian, symmetry breaking, and particle mass generation relevant to the subsequent analysis. In Section 3, we derive analytic expressions for the half-life and the effective Majorana mass, incorporating both the standard and new physics mechanisms, followed by numerical calculations and a comprehensive discussion. Finally, Section 4 summarizes the main conclusions.

\section{\label{model} A review of the model}
 
\subsection{Symmetry and field content}

The gauge symmetry of this flipped trinification and $SU(3)_L,SU(3)_R$ interchange preserved model is derived by left-right symmetrizing the 3-3-1 group \cite{Singer:1980sw, Pisano:1992bxx, Frampton:1992wt, Foot:1992rh, Montero:1992jk, Dong:2017zxo, Dinh:2019jdg}
, which can be simply expressed as
\be SU(3)_C\otimes SU(3)_L\otimes SU(3)_R \otimes U(1)_X,\ee
where $SU(3)_C$ is the usual QCD gauge group, while $SU(3)_L$ and $SU(3)_R$ are, respectively, the gauge groups of the left and right sectors. $U(1)_X$ is a given $U(1)$ symmetry before the gauge symmetry is spontaneously 
broken, thus it is not the exact gauge group of the QED. The electric charge operator, then, is given by  
\be
Q=T_{3L}+T_{3R}+\beta\left(T_{8L}+T_{8R}\right)+X,
\ee where $T_{iL,R}$ $(i=1,2,3,...,8)$ and $X$ are the generators of $SU(3)_{L,R}$ and $U(1)_X$, respectively. The non-commutative baryon minus lepton number is identified as   
\be 
\frac{1}{2}\left(B-L \right)=\beta\left(T_{8L}+T_{8R}\right)+X,
\ee
which is, thus, in contrast to the usual (Abelian) extensions. The numerical constant $\beta$ associates with a basic electric charge defined as $q=-(1+\sqrt{3}\beta)/2$. In this work, we are interested in the scenario of model with three dark matter candidates in which $q=0$ thus $\beta=-1/\sqrt{3}$ \cite{Dong:2017zxo, Dinh:2019jdg}.\\

In analogy, the fermion content is derived from that of the 3-3-1 model through the application of left–right symmetrization, which leads to
  \be \psi_{aL} =
\left(\begin{array}{c}
               \nu_{aL}\\ e_{aL}\\ N^q_{aL}
\end{array}\right) \sim \left(1,3, 1,\frac{q-1}{3}\right),  \hs \hs  \psi_{aR} =
\left(\begin{array}{c}
               \nu_{aR}\\ e_{aR}\\ N^q_{aR}
\end{array}\right) \sim \left(1,1, 3,\frac{q-1}{3}\right),\ee
\be Q_{\alpha L}=\left(\begin{array}{c}
  d_{\alpha L}\\  -u_{\alpha L}\\  J^{-q-\frac{1}{3}}_{\alpha L}
\end{array}\right)\sim \left(3,3^*,1,-\frac{q}{3}\right), \hs \hs
Q_{\alpha R}=\left(\begin{array}{c}
  d_{\alpha R}\\  -u_{\alpha R}\\  J^{-q-\frac{1}{3}}_{\alpha R}
\end{array}\right)\sim \left(3,1,3^*,-\frac{q}{3}\right), \ee   \be  Q_{3L}= \left(\begin{array}{c} u_{3L}\\  d_{3L}\\ J^{q+\frac{2}{3}}_{3L} \end{array}\right)\sim
 \left(3,3,1,\frac{q+1}{3}\right), \hs \hs  Q_{3R}= \left(\begin{array}{c} u_{3R}\\  d_{3R}\\ J^{q+\frac{2}{3}}_{3R} \end{array}\right)\sim
 \left(3,1,3,\frac{q+1}{3}\right).\ee
Here, $a=1,2,3$ and $\alpha=1,2$ are generation indices, which are usually used in the previous researches \cite{Dong:2017zxo, Dinh:2019jdg}. Anomaly cancellation is independently guaranteed in both the left and right sectors owing to the equality between multiplets and anti-multiplets. The left–right symmetric structure of the model is preserved by the introduction of additional fermions, 
$N_a,$ and $ J_a$, in addition to the right-handed neutrinos $\nu_{aR}$.\\       

The scalar sector is also introduced with left-right symmetry. There are five scalar multiplets out there in the model, 
which are a $(3,3^*)$ multiplets, two L(R) triplets and two L(R) sextets. The detail is following:   
\be
\phi = \left(
\begin{array}{ccc}
 \phi_{ 11}^0 & \phi_{ 12}^+& \phi_{ 13}^{-q} \\
  \phi_{21}^- & \phi_{ 22}^{0} & \phi_{ 23}^{-1-q} \\
  \phi_{ 31}^{q}& \phi_{ 32}^{1+q}& \phi_{ 33}^{0} \\
\end{array}
\right)     \sim (1,3,3^*,0), \crn
\ee
\be
 \chi_L = \left( \begin{array}{ccc}\chi_1^{-q}\\ \chi_2^{-q-1}\\ \chi_3^0 \end{array} \right)_L \sim\left(1,3,1,-\frac{2q+1}{3} \right), \crn
\ee
\be
 \chi_R = \left( \begin{array}{ccc}\chi_1^{-q}\\ \chi_2^{-q-1}\\ \chi_3^0 \end{array} \right)_R \sim\left(1,1,3,-\frac{2q+1}{3} \right), 
 \label{HiggsSector}
 \ee
 \be
 \sigma_L = \left(%
\begin{array}{ccc}
 \sigma_{11}^0 & \frac{\sigma_{12} ^-}{\sqrt{2}}& \frac{\sigma_{ 13}^{q}}{\sqrt{2}} \\
  \frac{\sigma_{12}^-}{\sqrt{2}} & \sigma_{22}^{--} &\frac{ \sigma_{23}^{q-1}}{\sqrt{2}} \\
 \frac{ \sigma_{13}^{q}}{\sqrt{2}}&\frac{ \sigma_{23}^{q-1}}{\sqrt{2}}& \sigma_{33}^{2q} \\
\end{array}%
\right)_L     \sim \left(1,6,1,\frac{2(q-1)}{3}\right), \crn
\ee
\be
\sigma_R = \left(%
\begin{array}{ccc}
 \sigma_{11}^0 & \frac{\sigma_{12} ^-}{\sqrt{2}}& \frac{\sigma_{ 13}^{q}}{\sqrt{2}} \\
  \frac{\sigma_{12}^-}{\sqrt{2}} & \sigma_{22}^{--} &\frac{ \sigma_{23}^{q-1}}{\sqrt{2}} \\
 \frac{ \sigma_{13}^{q}}{\sqrt{2}}&\frac{ \sigma_{23}^{q-1}}{\sqrt{2}}& \sigma_{33}^{2q} \\
\end{array}%
\right)_R     \sim \left(1,1,6,\frac{2(q-1)}{3}\right).\crn\ee
Their corresponding expectation values (VEVs) implied from (\ref{HiggsSector}) are simply written as: 
\bea
\langle\phi \rangle&=&\frac{1}{\sqrt{2}} \left(%
\begin{array}{ccc}
u & 0 &0\\
 0 & u^\prime &0 \\
0& 0&w \\
\end{array}%
\right), \\  
\langle \chi_L \rangle &=&\frac{1}{\sqrt{2}} \left(%
\begin{array}{ccc}
	0\\
	0 \\
	w_L \\
\end{array}%
\right),\hs \langle \chi_R \rangle =\frac{1}{\sqrt{2}} \left(%
\begin{array}{ccc}
0\\
 0 \\
w_R \\
\end{array}%
\right), \label{vev2} \\ 
\langle \sigma_L \rangle &=&\frac{1}{\sqrt{2}} \left(%
\begin{array}{ccc}
	\Lambda_L & 0 &0\\
	0 & 0&0 \\
	0& 0&0 \\
\end{array}%
\right),\hs \langle \sigma_R \rangle =\frac{1}{\sqrt{2}} \left(%
\begin{array}{ccc}
\Lambda_R & 0 &0\\
 0 & 0&0 \\
0& 0&0 \\
\end{array}%
\right). \label{vev1}
  \eea   
  
The work in \cite{Dong:2017zxo} demonstrates that the pattern of symmetry breaking is not unique; instead, it can proceed through several possible sequences, each determined by how the vacuum expectation values (VEVs) are ordered in magnitude. These different hierarchies dictate which subgroup is broken at each stage. Remarkably, even though the breaking chains may differ, every viable scheme leaves behind a residual discrete gauge symmetry that remains unbroken by all VEVs. This surviving symmetry $W_P$ - named as matter parity - plays an important role because it constrains the allowed interactions and ensures the stability of certain fields, which are favour for dark matter candidates. The residual symmetry is expressed as
\be 
W_P=(-1)^{3(B-L)+2s}=(-1)^{6[\beta(T_{8L}+T_{8R})+X]+2s}.
\ee 
We work on the scheme $\Lambda_R, w_R, w \gg u,u^\prime \gg \Lambda_L, w_L$, which entirely appropriates to the potential minimization. Detail calculation shows that the minimization conditions imply $\Lambda_L\simeq 0,\ w_L\simeq 0$, where the small nonzero values come from abnormal perturbative interactions. In the chosen VEV hierarchy, the gauge symmetry is broken sequentially, first down to the Standard Model gauge group supplemented by matter parity, and subsequently to the residual symmetry
$SU(3)_C\otimes U(1)_Q\otimes W_P$. Note that the left–right asymmetry can be obviously manifested at the electroweak scale due to the vacuum configuration satisfies to $w\neq 0$, $w_R\neq w_L$ and $\Lambda_R\neq \Lambda_L$. 
\subsection{Mass generation}
In this section, we undertake a systematic examination of the physical states and mass spectra of the particles incorporated within the model under consideration. Our analysis is restricted to those particles that play a direct role in the phenomenology of neutrinoless double beta decay, which will be addressed in detail in a subsequent section. The mechanisms responsible for mass generation are presented in a stepwise manner, encompassing fermions, gauge bosons, and Higgs scalars.   
\subsubsection{Fermion masses}
The fermion masses originate from Yukawa interactions, which can be explicitly expressed as
\bea \mathcal{L}_{\mathrm{Yukawa}} & =& x_{ab}\left( \bar{\psi}^c_{aR} \sigma^\dagger_R \psi_{bR}+\bar{\psi}^c_{aL} \sigma^\dagger_L \psi_{bL}\right) + y_{ab} \bar{\psi}_{aL} \phi \psi_{b R}+\frac{z_{ab}}{M}\bar{\psi}_{aL}\chi_L \chi_R^* \psi_{bR} \crn
&&+ k_{33}\bar{Q}_{3L} \phi Q_{3R} +k_{\alpha \beta}\bar{Q}_{\alpha L} \phi^* Q_{\beta
R}+ \frac{k'_{33}}{M}\bar{Q}_{3L} \chi_L \chi_R^* Q_{3 R}+\frac{k'_{\alpha \beta}}{M} \bar{Q}_{\alpha L}\chi_L^* \chi_R Q_{\beta R}  \nonumber\\ 
&& + \frac{t_{3\alpha}}{M} \left(\bar{Q}_{3L}\phi \chi_R^* Q_{\alpha R}+\bar{Q}_{3R}\phi^* \chi_L^* Q_{\alpha L}\right) + \frac{t_{\alpha 3}}{M} \left(\bar{Q}_{\alpha L} \phi^* \chi_R Q_{3R}+\bar{Q}_{\alpha R}\phi \chi_{L}Q_{3L} \right)\crn
&& +H.c.,  \label{yukawa}\eea where $M$ denotes the characteristic scale of new physics governing the effective interactions. The imposition of left–right symmetry requires that the couplings $y,z,k,k'$ be Hermitian, while $x$ and $t$ remain unconstrained and may be taken as generic.\\

Fermion masses are generated from the Yukawa Lagrangian (\ref{yukawa}) after the symmetry spontaneously breaking. As expected, the standard model charged leptons obtain their masses at the electroweak scale,
\bea
\mathcal{L}_{\mathrm{mass}}^l= \left(\frac{y_{ab}}{\sqrt{2}}u^\prime \right) \bar{l}_{aL} l_{bR}+ H.c.
\label{mal}\eea 

\noindent The newly introduced leptonic states acquire a substantial mass at the emergent physical scale, expressed as
\bea \mathcal{L}_{\mathrm{mass}}^N = \left(\frac{y_{ab}}{\sqrt{2}}w+\frac{z_{ab}}{2M}w_L w_R\right) \bar{N}_{aL} N_{bR} +H.c.
\label{maN}\eea
It is important to emphasize that these novel leptons remain unmixed with the standard model leptons as a consequence of matter parity conservation. Furthermore, in the absence of effective interaction terms (see eqs. (\ref{mal}) and (\ref{maN})), their mixing matrices coincide with those of the ordinary leptons.\\ 

Mass matrix for neutrinos obtained from the Lagrangian (\ref{yukawa}) sustains the complete form of the type I seesaw with both kinds of Dirac and Majorana mass terms. In the basis $(\nu_L, \nu^c_R)$, its explicit expression is commonly given by 
\bea
\mathcal{M}_\nu =\left(%
\begin{array}{cc}
	M_\nu^L & M_\nu^D \\
	(M_\nu^D)^T & M_\nu^R \\
\end{array}%
\right),
\eea
where $M_\nu^D, M_\nu^L$ and $M_\nu^R$ are directly implied from (\ref{yukawa}) as
\bea
\left(M_\nu^D \right)_{ab} =-\frac{y_{ab}}{\sqrt{2}} u, \hs \left(M_\nu^L \right)_{ab}=-\sqrt{2}x_{ab}\Lambda_L, \hs
\left(M_\nu^R \right)_{ab}=-\sqrt{2}x_{ab}\Lambda_R.
\eea 
Given the hierarchical condition $\Lambda_L\ll u\ll \Lambda_R$, the active neutrinos ($\sim \nu_L$) acquire small masses through the seesaw mechanism, 
\be M_\nu\simeq -\sqrt{2} x \Lambda_L+\frac{1}{2\sqrt{2}}yx^{-1}y^T\frac{u^2}{\Lambda_R},\ee In contrast, the sterile neutrinos ($\sim \nu_R$) obtain large masses at the $\Lambda_R$ scale, characterized by $M'_\nu\simeq -\sqrt{2}x\Lambda_R$.\\   

In the context of quark content, detailed considerations presented in \cite{Dong:2017zxo} demonstrate that the ordinary quarks acquire masses consistent with current experimental observations at the electroweak scale, of the order $u$, $u'$. In contrast, the newly introduced quarks are characterized by significantly heavier masses, which arise at the scale associated with new physics beyond the standard model. Since these results are not directly relevant to the subsequent analysis, we refrain from presenting them in detail here.

\subsubsection{Gauge boson masses \label{gaugemass}}
Note that, in the present model, the gauge boson content and their associated mass spectrum exhibit only minor deviations from those reported in \cite{Dong:2017zxo}, primarily due to the small vacuum expectation values of $\sigma_L$ and $\chi_L$. Hereafter, we restrict ourselves to a concise summary of the principal results-comprehensively detailed in \cite{Dinh:2019jdg}-and focus specifically on the singly charged gauge bosons, as these play a central role in mediating the neutrinoless double beta decay process beside the doubly charged Higgs scalars.\\

The gauge sector of the model comprises a total of eleven bosons: two singly charged, two carrying charge $q$, two carrying charge $q+1$, and five neutral states. Among the singly charged bosons, one acquires an electroweak-scale mass and is identified with the standard model boson ($W_1$), while the second ($W_2$) obtains a mass of order $\Lambda_R$. The additional gauge bosons with charges $q$ ($X_{1,2}^q$) and $q+1$ ($Y_{1,2}^{q+1}$) are heavy, with their masses determined by the parameters  $w$ and $w_R$. Of the five neutral gauge bosons, one remains massless and corresponds to the photon, while another acquires an electroweak-scale mass and is identified with the standard model $Z$ boson. The remaining three neutral states are heavy, with masses situated at the scale of new physics.\\

As our main interest, the physical singly charged gauge bosons, denoted as $W_1$ and, $W_2$ arise as linear combinations of the gauge eigenstates $W_L$ and $W_R$. The explicit expression is 
\be W_{1 }= c_\xi W_{L } -s_\xi W_{R },\hs W_{2 }= s_\xi W_{L } +c_\xi W_{R },\ee where the mixing angle $\xi$ is defined by \be t_{2 \xi} = \frac{4 t_R u u'}{2  \Lambda _L^2-2 \Lambda _R^2 t_R^2-\left(t_R^2-1\right) (u^2+ u'^2)}\simeq -\frac{2uu'}{t_R\Lambda^2_R}\ll 1.\ee 
Their masses are simply written as
\bea
 m_{W_1}^2 && \simeq \frac{g_L^2}{4}\left[u^2+u^{\prime 2}+2\Lambda_L^2-\frac{4t_R^2 u^2 u^{\prime 2}}{(t_R^2-1) (u^2+u^{\prime 2}) +2t_R^2 \Lambda^2_R-2\Lambda_L^2} \right]\simeq \frac{g^2_L}{4}(u^2+u'^2), \\ m_{W_2}^2  && \simeq \frac{g_R^2}{4} \left[u^2+u^{\prime 2}+2\Lambda_R^2+\frac{4 t^2_R u^2 u^{\prime 2}}{(t_R^2-1) (u^2+u^{\prime 2}) +2t_R^2 \Lambda^2_R-2\Lambda_L^2} \right]\simeq \frac{g^2_R}{2}\Lambda^2_R, \eea 
where $g_L,g_R$ are gauge couplings associated with the $SU(3)_{L}$ and $SU(3)_{R}$ symmetries, respectively. At the flipped trinification scale, the left–right symmetry enforces the relation
$t_R\equiv g_R/g_L=1$. However, at low energies this equality need not hold, since the renormalization group evolution of the couplings introduces distinct contributions to their running, leading to $t_R\neq 1$. Among the resulting gauge bosons, $W_1$ coincides with the standard model  $W$ boson. Its mass relation requires that the vacuum expectation values satisfy $u^2+u'^2=(246\ \mathrm{GeV})^2$ consistent with the electroweak scale. In contrast, $W_2$  represents a new singly charged gauge boson beyond the standard model, whose mass emerges from the extended symmetry breaking pattern.
  
\subsubsection{Higgs masses}
A comprehensive analysis of the scalar potential, incorporating the full scalar field content, has been carried out in \cite{Dinh:2019jdg}. Below, we summarize the relevant results. For scalar fields carrying charges $q+1$ and $+1$, the construction yields a $4\times 4$ mass matrix for each class of particles. It is important to recall that the model accommodates two massive gauge bosons associated with each type of charged state. To gain their massess, these gauge bosons have absorbed two massless Goldstone bosons—specifically, two with charge $q+1$ and two with charge $+1$. Consequently, the spectrum retains two heavy massive scalar states for each charge sector.\\

A potentially significant contribution to neutrinoless double beta decay arises through the exchange of doubly-charged scalar fields. The model predicts the existence of two heavy doubly-charged scalars, denoted as  $H_1^{\pm \pm}$ and $H_2^{\pm \pm}$. Their physical states are defined by the following mixing relations:
\bea
H^{\pm \pm}_{1}= c_{\xi_7}\sigma_{22R}^{\pm \pm}-s_{\xi_7}\sigma_{22L}^{\pm \pm}, \hs \hs H^{\pm \pm}_{2}&=& s_{\xi_7}\sigma_{22R}^{\pm \pm}+c_{\xi_7}\sigma_{22L}^{\pm \pm},
\eea
where the mixing angle  is determined by $t_{2\xi_7}=\frac{2u^{\prime 2}\zeta_9 \Lambda_L \Lambda_R}{(\Lambda_L^2-\Lambda_R^2)(-u^2\zeta_9+2\Lambda_L\Lambda_R)}$.
In the phenomenologically relevant limit $\Lambda_R \gg \Lambda_L$, the mixing angle approaches $\xi_7 \simeq 0$. Consequently, the fields  $\sigma_{22L}$, and $\sigma_{22R}$  effectively serve as physical states without significant mixing.\\ 

The spectrum of neutral scalar fields in the model can be classified into two distinct components: CP-even and CP-odd states. Within the CP-even sector, the theory predicts the existence of a single light neutral scalar, which is identified with the standard model Higgs boson. All additional CP-even states acquire masses at a new, higher physical scale, thereby decoupling from low-energy phenomenology. In contrast, the CP-odd sector exhibits a richer structure. It contains four massless Goldstone bosons, which are absorbed through the Higgs mechanism to provide the longitudinal degrees of freedom for four massive gauge bosons, namely $Z,Z_R, Z_L^\prime,$ and $Z_R^\prime$. Beyond these Goldstone modes, the CP-odd sector also accommodates three heavy scalar states, whose masses are generated at the same new physical scale as the heavy CP-even states.

\section{Phenomenology of the neutrinoless double beta decay}

\mathversion{bold}
\subsection{Neutrinoless double beta decay}
\mathversion{normal}

In the framework of flipped trinification, characterized by the gauge symmetry $SU(3)_C\otimes SU(3)_L\otimes SU(3)_R\otimes U(1)_X$, neutrinoless double beta decay can potentially occur at tree level through the exchange of neutrinos—both light and heavy—as well as doubly charged Higgs scalars. The corresponding interaction vertices relevant to this process are listed in detail in Table \ref{TableVertex}.\\

To have the Table \ref{TableVertex}, we have used notation that the correspondence between the gauge eigenstates and the mass eigenstates of the ordinary charged leptons ($l$) is formally expressed as
\bea
e_{a L}=(U^{l}_L)_{a k}e_{kL}^{\prime}, \hs e_{a R}=(U^{l}_R)_{a k}e_{kR}^{\prime},
\eea
where the unitary transformations $U^{l}_{L,R}$ denote the basis-changing (mixing) matrices. For completeness, we note that the mixing structures associated with the charged gauge bosons $Y^{\pm(q+1)}$ and the scalar fields $\mathcal{H}^{\pm(q+1)}$ are not explicitly presented here, as these states do not play a role in the neutrinoless double beta ($0\nu\beta\beta$) decay mechanism under consideration.\\

The neutrino mixing matrix is denoted by $U^\nu$, a $6\times 6$ unitary matrix that maps the gauge eigenstate basis, $X_L\equiv(\nu_L, (\nu_R)^c)^T$, to the corresponding mass eigenstate basis, $X'_L\equiv(\chi_{\nu L}, \chi_{N L})^T$, via the transformation relation $X_L=U^\nu X'_L$.
The explicit block structure of $U^\nu$ can be technically written as
\bea
U^\nu=\left(
  \begin{array}{cc}
      U_L & U_A \\
      U_B& U_R  \\
  \end{array}
\right)=\left(
  \begin{array}{c}
      U_L^\nu \\
      U_R^\nu \\
  \end{array}
\right).
\eea
where $U_{L,R,A,B}$ denote the submatrices encoding the mixing between light and heavy neutrino sectors. In this framework, the Yukawa coupling $x$ admits the compact representation
\be
\label{YukawaX}
x=-\frac{\left(U_L^*m_LU_L^\dagger+U_A^*m_RU_A^\dagger\right)}{\sqrt2\Lambda_L},
\ee
with $m_L$ and $m_R$ denoting the diagonal mass matrices of the light and heavy neutrino states, respectively.
Analogously, the Yukawa coupling $y$ can be expressed in terms of the diagonal mass matrices $m_l$ of the charged leptons, together with the mixing matrices $U_{L,R}^{l}$:
\be
\label{YukawaY}
y=-\frac{\sqrt2U_L^l m_l\left(U_R^l\right)^\dagger}{u'}.
\ee 
It is useful to remind that, beside the charged lepton mass $m_l$, neutrino Dirac mass matrix is also proportional to $y$, $m_\nu^D\sim y u$.
\\
\begin{table}[h]
\begin{center}
	\scalebox{1}{
		\begin{tabular}{|c|c|}
			\hline
			Vertex & Coupling \\
			\hline
			$\left(\bar{e}'_{L}\gamma^\mu \chi_{\nu L}\right) W_{i\mu}^{-}$ &$\frac{-ig_L}{\sqrt{2}}U_L^{W_{i\mu}^{-}}=\frac{-ig_L}{\sqrt{2}}(U^{l\dagger}_L U_{L})c_{\xi}~(s_{\xi})  $ \\
			\hline
			$\left(\bar{e}'_{L}\gamma^\mu \chi_{N L}\right) W_{i\mu}^{-}$ &$\frac{-ig_L}{\sqrt{2}}V_L^{W_{i\mu}^{-}}=\frac{-ig_L}{\sqrt{2}}(U^{l\dagger}_L U_A)c_{\xi}~(s_{\xi})  $ \\
			\hline
			$\left(\bar{e}'_{R}\gamma^\mu \chi_{\nu R}\right) W_{i\mu}^{-}$ &$\frac{-ig_R}{\sqrt{2}}V_R^{W_{i\mu}^{-}}=\frac{-ig_R }{\sqrt{2}}(U^{l\dagger}_R U_B)s_{\xi}~(c_{\xi})  $ \\	
			\hline
			$\left(\bar{e}'_{R}\gamma^\mu \chi_{N R}\right) W_{i\mu}^{-}$ &$\frac{-ig_R}{\sqrt{2}}U_R^{W_{i\mu}^{-}}=\frac{-ig_R }{\sqrt{2}}(U^{l\dagger}_R U_{R})s_{\xi}~(c_{\xi})  $ \\			
			\hline
			$\bar{e}'_{L} e'^c_{L} H_{i}^{--}$ &$ Y_{H_i^{--}}^L= -i(U^l_L)^\dagger x (U^{l}_L)^*c_{\xi_6}~(s_{\xi_6}) $\\ 			
			\hline
			$\bar{e}'_{R} e'^c_{R} H_{i}^{--}$ &$ Y_{H_i^{--}}^R=-i(U^l_R)^\dagger x (U^{l}_R)^*s_{\xi_6}~(c_{\xi_6}) $\\ 			
			\hline
	\end{tabular}}
	\caption{Vertices involving in the process of neutrinoless double beta decay. }
	\label{TableVertex}
	\end{center}
\end{table}
\begin{figure}
\begin{center}
\begin{tabular}{ccc}
\includegraphics[width=5cm,height=3.5cm]{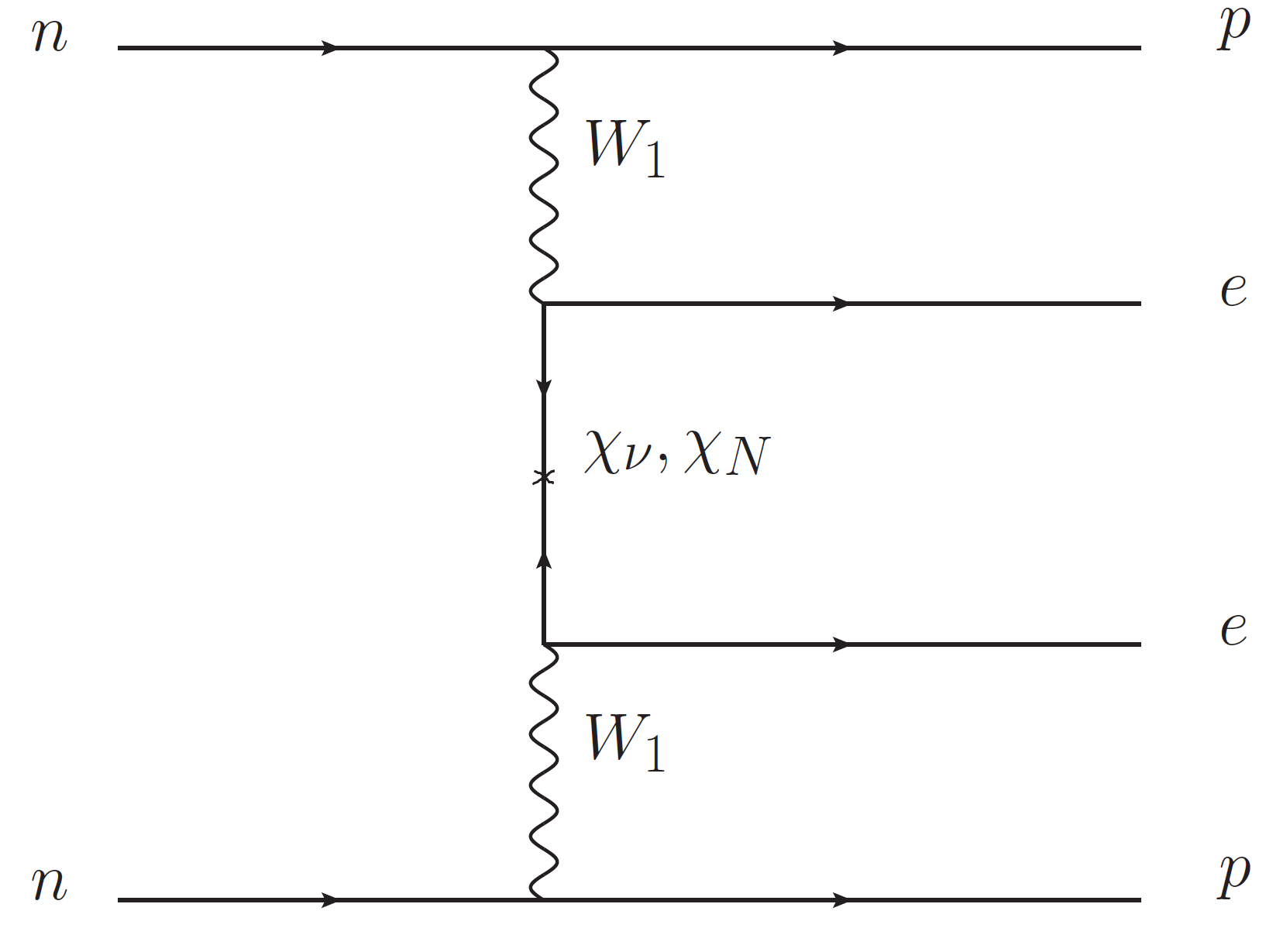}&
\includegraphics[width=5cm,height=3.5cm]{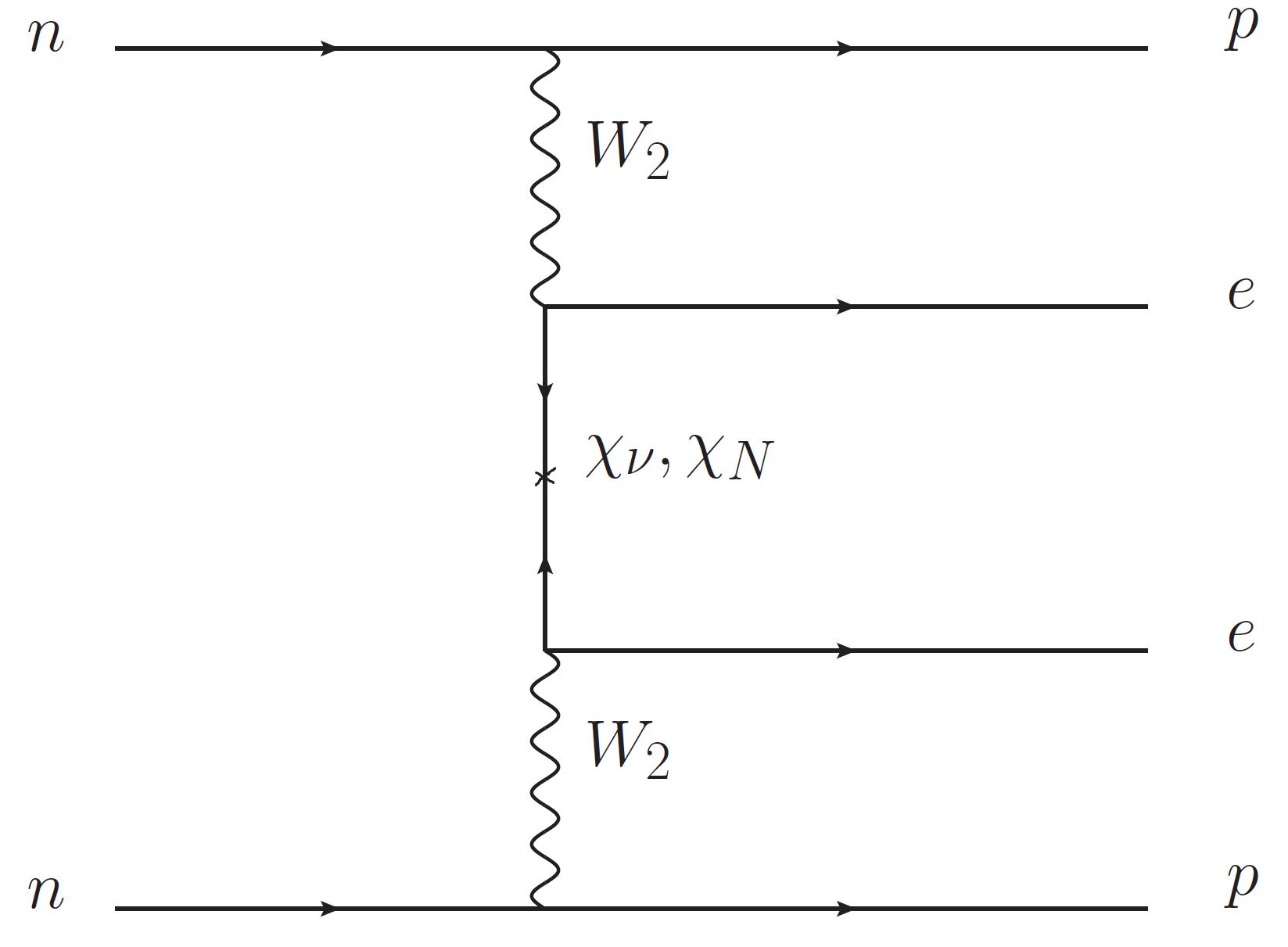}&
\includegraphics[width=5cm,height=3.5cm]{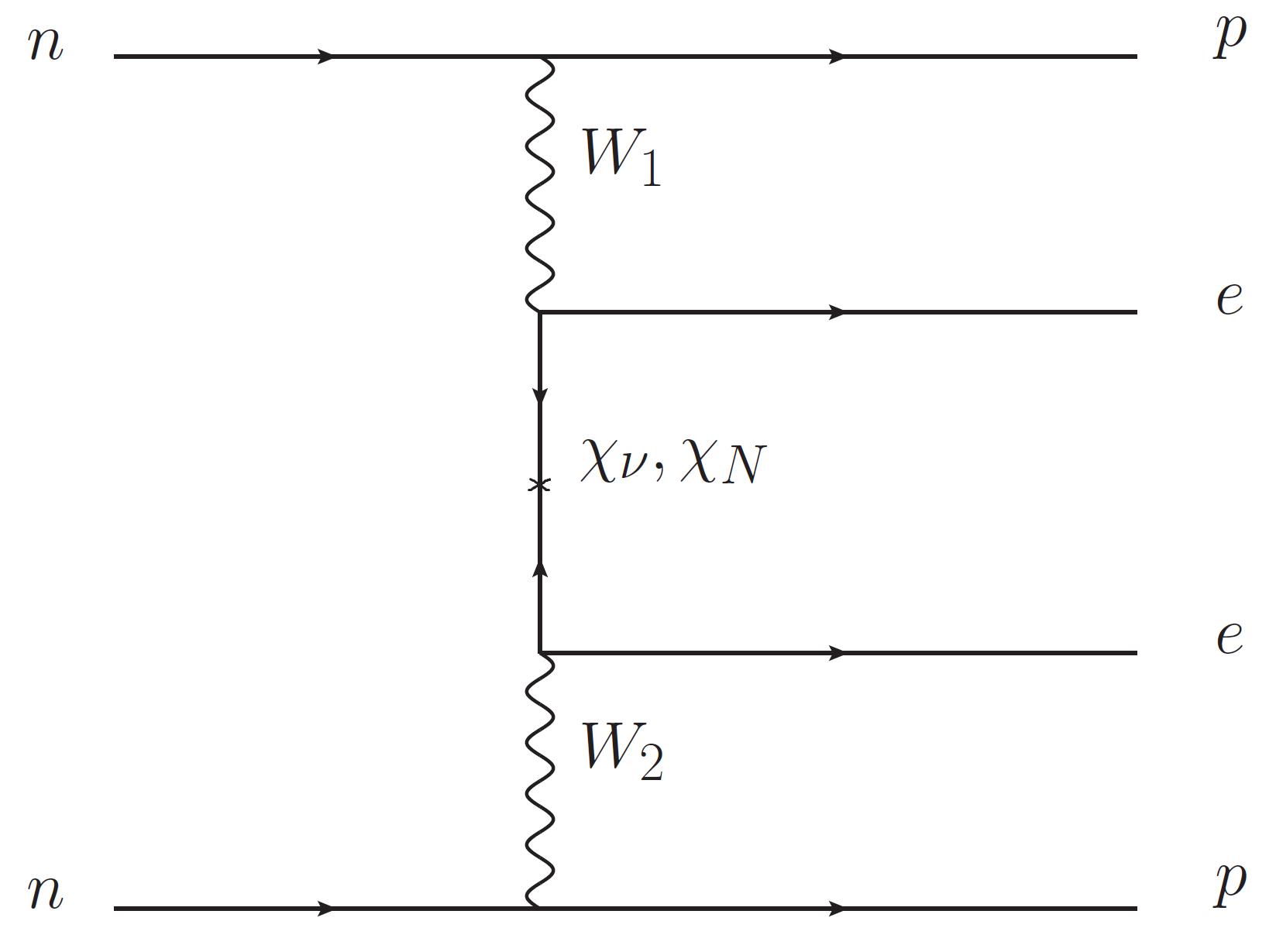}
\end{tabular}
\caption{Neutrinoless double beta decay contribution from the light and heavy
 Majorana neutrino exchanges, mediated by singly-charged gauge bosons.}
\label{GaugeExc}
\end{center}
\end{figure}

\begin{figure}
\begin{center}
\begin{tabular}{cc}
\includegraphics[width=5cm,height=3.5cm]{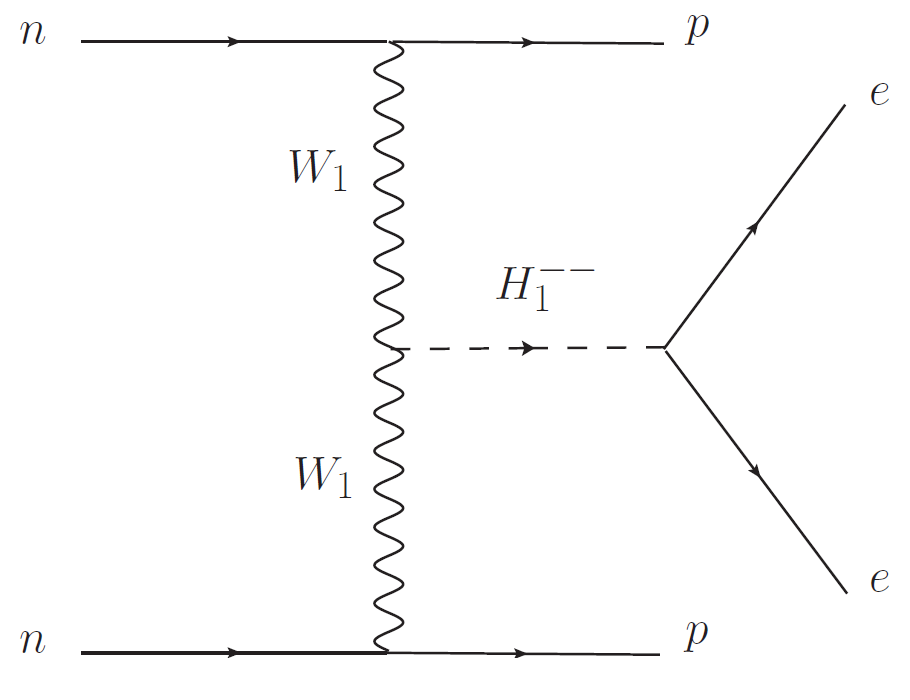}&
\includegraphics[width=5cm,height=3.5cm]{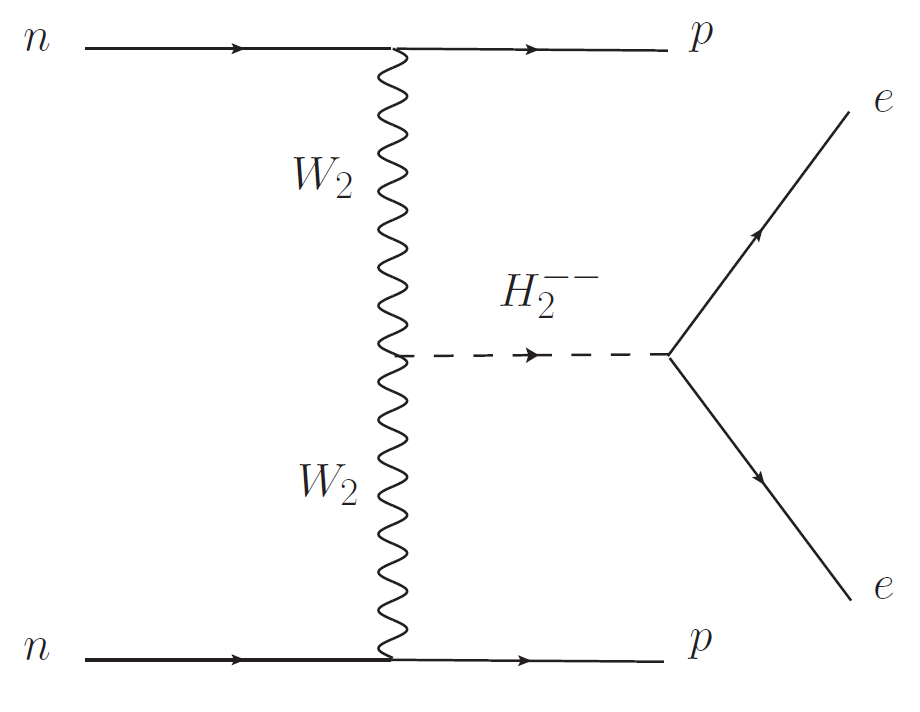}
\end{tabular}
\caption{Neutrinoless double beta decay contribution provided by the channels of doubly-charged Higgs scalar mediation.}
\label{HiggExc}
\end{center}
\end{figure}
Thus, in the considered model, the dominant contributions to the rate of $0\nu\beta\beta$ decay process are provided by the standard mechanism due to the exchange of light neutrino with mediation of $W_1$ gauge boson, and the new contributions due to the exchanges of heavy neutrinos and doubly charged Higgs scalars. The Feynman diagrams for the main contributions to the $0\nu\beta\beta$ decay amplitude are shown in the Fig. \ref{GaugeExc} and Fig. \ref{HiggExc}.
The inverse half-life for the process can be written as \cite{Chakrabortty:2012mh, Banerjee:2023aro, BhupalDev:2014qbx, Patra:2023ltl, Simkovic:1999re, Simkovic:2009pp, Faessler:1998sba}
\begin{equation}
\left(T^{\nonubb}_{1/2}\right)^{-1}= g_A^4{G_{01}^{0\nu}} 
\Big|
\mathcal{M}_{\nu_L}^{0\nu}\Big( \eta^{LL}_{\nu_L}  + \eta^{RR}_{\nu_L} + \eta^{LR}_{\nu_L} 
+\eta_{H_{1}}\Big)    
+ ~\mathcal{M}_{\nu_R}^{0\nu} \Big(\eta^{LL}_{\nu_R}   +   
\eta^{RR}_{\nu_R}   +   
\eta^{RL}_{\nu_R}    
+ \eta_{H_2}\Big)
\Big |^2    ,
\label{Lifetime} 
\end{equation} 
where $g_A$ is the axial coupling constant, $G_{01}^{0\nu}$ is the phase space factor, $\mathcal{M}_\nu^{0\nu}~(\mathcal{M}_{\nu_R}^{0\nu})$ are the nuclear matrix elements (NME) for long-distance (short-distance) interactions, which are detail provided in Table \ref{tab:nucl-matrix}, and the dimensionless parameters $\eta$ are defined as follows: 
\\
$$
\mathcal{\eta}^{LL}_{\nu_L} =  \frac{\left(U_L^{W_{1\mu}}\right)^2_{e\,i}\, m_i}{m_e}; \nonumber \quad 
\mathcal{\eta}^{LL}_{\nu_R} =  \frac{\left(V_L^{W_{1\mu}}\right)^2_{e\,i} m_p}{M_i} ;\nonumber 
$$
\\
$$\mathcal{\eta}^{RR}_{\nu_R} = \frac{M^4_{W_1}}{M^4_{W_2}} \frac{\left(U_R^{W_{2\mu}}\right)^2_{e\,i} m_p}{M_i}; \nonumber
\quad
\mathcal{\eta}^{RR}_{\nu_L} = \frac{M^4_{W_1}}{M^4_{W_2}} \frac{\left(V_R^{W_{2\mu}}\right)^2_{e\,i}\, m_i}{m_e} ;
$$\\
$$
\mathcal{\eta}^{LR}_{\nu_L} = \frac{M^2_{W_1}}{M^2_{W_2}} \frac{\left(U_L^{W_{1\mu}}\right)_{e\,i} \left(V_R^{W_{2\mu}}\right)_{e\,i}\, m_i}{m_e}; \nonumber
\quad
\mathcal{\eta}^{RL}_{\nu_R} = \frac{M^2_{W_1}}{M^2_{W_2}} \frac{\left(U_R^{W_{2\mu}}\right)_{e\,i} \left(V_L^{W_{1\mu}}\right)_{e\,i}\, m_p }{M_i}; \nonumber
$$
\\
\be \eta^{}_{H_1} = \frac{Y_{H_1}^L\Lambda_Lm_p}{M_{H_1}^2};
~~
\eta^{}_{H_2} = \frac{M^4_{W_1}}{M^4_{W_2}} \frac{Y_{H_2}^R\Lambda_R m_p}{M_{H_2}^2}.
\label{etadef}
\ee
Let us rewrite equation (\ref{Lifetime}) in form
\be
\left(T^{\nonubb}_{1/2}\right)^{-1}= g_A^4{G_{01}^{0\nu}}
\Big|\frac{\mathcal{M}_{\nu_L}^{0\nu}}{m_e}
\Big |^2
\Big|m^{eff}_{\beta\beta},
\Big |^2    ,
\label{Lifetime1} 
\ee
in which $m^{eff}_{\beta\beta}$ is called as the effective Majorana mass. Current upper limits on the effective mass of the neutrino less double beta decay for different atoms (isotopes) are shown in the Table \ref{tab:nucl-matrix}. 
\begin{table}[h]
 \centering
\vspace{10pt}
 \begin{tabular}{|l|cc|cc|cc|cc|c|}
 \hline \hline
  &  $^{76}$Ge 
& & $^{82}$Se  & & $^{130}$Te & &$^{136}$Xe &  \\[1mm] 
{methods}  & {${\cal M}^{0\nu}_{\nu_L}$}  
 & {${\cal M}^{0\nu}_{\nu_L}$} & {${\cal M}^{0\nu}_{\nu_L}$} & {${\cal M}^{0\nu}_{\nu_R}$}&{${\cal M}^{0\nu}_{\nu_L}$} &{${\cal M}^{0\nu}_{\nu_R}$} &{${\cal M}^{0\nu}_{\nu_L}$}  & {${\cal M}^{0\nu}_{\nu_R}$} \\[1mm]
 \hline 
 dQRPA \cite{Fang:2018tui} & $3.12$  & $187.3$ & $2.86$  & $175.9$& $2.90$& $191.4$ &$1.11$ & $66.9$\\ 
  \hline
  QRPA-Tu \cite{Simkovic:2013qiy,Faessler:2014kka}  & $5.16$  & $287.0$ &$4.64$  & $262.0$&$ 3.89$ & $264.0$ & $2.18 $ & $152.0$ \\ 
  \hline
  QRPA-Jy \cite{Hyvarinen:2015bda}  & $5.26$  & $401.3$ & $3.73$ &$287.1$ &$4.00$ &$338.3$ & $2.91 $ & $186.3$ \\  
  \hline
   IBM-2 \cite{Barea:2013bz} & $4.68$  & $104$ & $3.73$ & $82.9$ & $3.70 $& $91.8 $ & $3.05 $ & $72.6 $ \\ 
  \hline
  CDFT \cite{Song:2014vra,Yao:2014uta,Song:2017ktj} & $ 6.04$ & $209.1 $  & $5.30 $ & $189.3 $ & $4.89 $ & $193.8 $ & $4.24 $ & $166.3 $ \\ 
\hline 
 ISM \cite{Menendez:2017fdf} & $2.89 $ & $130 $  & $2.73 $ & $121 $ & $2.76 $ & $146 $ & $ 2.28$ & $116 $ \\ 
\hline
\hline
$G^{0\nu}_{01}$ $[{10^{-14} \rm yrs}^{-1}] $ \cite{Horoi:2017gmj} & $0.22$ & & 1 & & $1.4$ & &$1.5$ &\\
\hline \hline
 \end{tabular}
 \caption{Phase space factor $G^{0\nu}_{01}$ and values of Nuclear Matrix Elements for various isotopes calculated 
by different methods for light and heavy neutrino exchange \cite{Horoi:2017gmj, Ejiri:2019ezh}.}
 \label{tab:nucl-matrix}
\end{table} 

\begin{table}[h]
\centering
	\begin{tabular}{cc|c|c}
	\hline
	Isotope & $T_{1/2}^{0 \nu}~\text{ yrs}$ & $m_{\beta \beta}^{0 \nu}~[\text{eV}]$ & Collaboration \\
	\hline
	$^{76}$Ge	& $> 1.8\times10^{26}$	& $< (0.08 - 0.18)$ 	& GERDA~\cite{GERDA:2020xhi} 			\\
	$^{76}$Ge	& $> 2.7\times10^{25}$	& $< (0.2 - 0.433)$ 	& MAJORANA DEMONSTRATOR~\cite{Majorana:2019nbd} 			\\
		& $> 8.3\times10^{25}$	& $< (0.113 - 0.269)$ 	& ~\cite{Majorana:2022udl} 			\\
	$^{82}$Se	& $> 3.5\times10^{24}$	& $< (0.311 - 0.638)$ 	& CUPID-0~\cite{CUPID:2019gpc} 			\\
	$^{130}$Te	& $> 2.2\times10^{25}$	& $< (0.09 - 0.305)$ 	& CUORE~\cite{CUORE:2021mvw} 			\\
	$^{136}$Xe	& $> 3.5 \times10^{25}$	& $< (0.093 - 0.286)$ 	& EXO~\cite{EXO-200:2019rkq} 				\\
	$^{136}$Xe	& $> 1.07\times10^{26}$	& $< (0.061 - 0.165)$ 				& KamLAND-Zen~\cite{KamLAND-Zen:2016pfg} 		\\
		& $> 2.3\times10^{26}$	& $< (0.036 - 0.156)$ 				& ~\cite{KamLAND-Zen:2022tow} 		\\
	\\
	\hline
	\end{tabular}	
\caption{The current lower limits on the half-life $T_{1/2}^{0 \nu}$ and their corresponding upper limits on 
the neutrinoless double beta decay effective mass $m_{\beta\beta}^{0 \nu}$  for different isotopes.}
\label{tab:halflife}
\end{table}
%
\subsection{Numerical analysis}
\label{SectionWL0}
\mathversion{normal}

Before proceeding with numerical evaluations based on the formula derived in the previous section, it is instructive to estimate the magnitudes of the relevant vacuum expectation values (VEVs). Among the introduced parameters, the smallest scale is expected to be $\Lambda_L$, which lies at the eV scale of neutrino masses. This value is significantly smaller than the weak-scale VEVs $u,u'$, which satisfy the electroweak constraint $u^2+u'^2=(246~{\rm GeV})^2$. Constraints from quark flavor-changing neutral currents (FCNCs) impose lower bounds on the scales $w,w_R$ and $\Lambda_R$, requiring them to be at least of order $w,w_R,\Lambda_R \gtrsim \mathcal{O}(50-100)$ TeV. These values are consistent with collider bounds \cite{Dong:2017zxo}, and correspond to the scales at which the flipped trinification symmetry is broken down to the Standard Model. Consequently, these VEVs are substantially larger than the weak scales. The parameter  can assume different hierarchies relative to $w$ and $w_R$, depending on the specific symmetry breaking scheme: i) $\Lambda_R\gg w,w_R$, (ii) 
$\Lambda_R\sim w,w_R$, or (iii) $\Lambda_R\ll w,w_R$. Phenomenologically viable dark matter scenarios \cite{Huong:2016ybt, Dong:2017zxo} favor cases (i) and (ii), which will be the focus of subsequent analysis.\\

Given the hierarchy of vacuum expectation values (VEVs) $\Lambda_R,w,w_R\gg u,u'\gg w_L,\Lambda_L$, the masses 
of the singly charged gauge bosons can be approximated as
$m^2_{W_1}\simeq \frac{g^2}{4}(u^2+u'^2)$, $m^2_{W_2}\simeq \frac{g^2}{2}\Lambda^2_R$
where the equality $g_L=g_R=g$ has been assumed. The lighter state $W_1$ acquires a mass identical to that of the Standard Model  boson, while the heavier state $W_2$ obtains a mass at the TeV scale or higher. Current collider bounds require $m_{W_2}\geq 5$ TeV, which translates into the condition $\Lambda_R\geq 10.8$ TeV.
The masses of the new scalar states, particularly the doubly charged Higgs bosons $H^{\pm\pm}$, depend on parameters in the scalar potential that are not precisely determined. Nevertheless, their masses are expected to scale with the new physics parameters $\Lambda_R,w,w_R$. To remain consistent with non-observation in collider searches \cite{Tanabashi:2018oca}, these masses must be sufficiently large. A reasonable choice places them in the range from several hundred GeV to a few TeV. In scenarios with hierarchical symmetry breaking, the upper bound can extend to values of order hundreds of TeV.\\

Without loss of generality, we adopt the basis of normal charged lepton mass eigenstates, in which the mass matrix $m_l=yu'/\sqrt 2$ is diagonal and the unitary transformations satisfy $U_{L,R}^l=I$, consistent with the standard model treatment. Moreover, the masses of the new leptons can be approximated as $m_N \simeq yw/\sqrt 2$ since the correction term $w_Lw_R/M\ll w$ is negligible. Consequently, the ratio of charged lepton to new lepton masses is basis-independent and universal across generations:
 \be
 \label{memE}
 \frac{m_l^i}{m_N^i}\simeq\frac{u'}{w}.
 \ee
The quantum numbers of the new lepton  depend on the parameter $\beta$. For $\beta=-1/\sqrt{3}$, $N$ is a Standard Model singlet, analogous to a light sterile neutrino, and this case remains phenomenologically viable. In contrast, for $\beta=1/\sqrt{3}$, $N$ carries electric charge $q=-1$. This scenario is strongly constrained by electroweak precision tests and is excluded unless the new physics scale is raised to an extreme hierarchy, $w/u'\gtrsim 10^5$ (see \cite{Dong:2017zxo, Dinh:2019jdg}).
For the neutral new lepton case ($\beta=-1/\sqrt{3}$), mixing effects allow the Higgs boson  to decay into light $N$ states. However, the decay rate is suppressed by factors of order $(u,u')^2/(w,w_R,\Lambda_R)^2\ll 1$. As a result, these light  states interact only very weakly and remain undetectable in current experiments. Their masses are therefore unconstrained and may take values consistent with the specific scenario under consideration.\\
  
In this study, we focus on models characterized by a new physics scale that is not excessively high, thereby allowing for a rich spectrum of physical phenomena accessible to current and forthcoming experimental programs. A key parameter in this framework is $u'$. To remain close to the Standard Model, we consider $u'$ near its maximal value, specifically $u'\simeq 246$ GeV. This choice has the advantage of suppressing the scale $u$, which in turn significantly reduces  $\Lambda_R$ (with $M_R\sim x\Lambda_R$), while still satisfying the seesaw condition $u^2/\Lambda_R\sim \rm{eV}$.
The observed neutrino masses require both $M_L$ and the seesaw contribution $M_D^T M_R^{-1}M_D$ to be of order eV. The smallness of $M_L$ is ensured by a correspondingly small $\Lambda_L\sim\mathrm{eV}$. Meanwhile, the magnitude of the seesaw term is given by $M_D^TM_R^{-1}M_D=\frac{u^2}{2}y^TM_{R}^{-1}y$, which depends on the interplay between the Yukawa couplings $y$, the scale $u$, and the right-handed neutrino mass $M_R$.
For the choice $u'\simeq 246$ GeV, the Yukawa couplings are fixed as $\mathrm{diag}(y)\simeq(3\times 10^{-6},6\times 10^{-4}, 10^{-2})$. Under these conditions, upper bounds on $u$ can be established: specifically, $u\leq 0.147$ GeV for $M_R\geq 1$ TeV, and $u\leq 0.445$ GeV for $M_R\geq 10$ TeV.\\

To summarize, the parametrization employed for the numerical analysis is as follows. The left-handed mixing matrix $U_L$ is identified with the Pontecorvo–Maki–Nakagawa–Sakata (PMNS) matrix, $U_{\mathrm{PMNS}}$, which has been determined with high precision from neutrino oscillation experiments, up to the uncertainties associated with the Dirac and Majorana CP-violating phases.
The right-handed mixing matrix $U_R$ is parametrized analogously to $U_{\mathrm{PMNS}}$, but in this case the mixing angles and CP phases are treated as free parameters in the calculation. The heavy neutrino mass matrix is then constructed via $M_R=U^*_R M_R^{diag} U^\dagger_R$,
where $M_R^{diag}$ denotes the diagonal matrix of heavy neutrino masses.
Finally, the Yukawa matrix $x$ is derived from $M_R$ according to the relation $x=-U^*_R M_R^{diag} U^\dagger_R/(\sqrt 2 \Lambda_R)$, where $diag(M_R^{diag})=(M_1,M_2,M_3)$.\\

For the vacuum expectation values (VEVs) in our framework, we adopt the following assignments. The parameter $\Lambda_L$ is chosen at the eV scale, while $u'$ is fixed at 246 GeV. Owing to the relation $u^2+u'^2=246^2~\rm{GeV^2}$, this choice implies that $u\sim 0.1$ GeV. Consequently, the right-handed neutrino mass matrix satisfy $M_R\geq 1$ TeV, as discussed above. Furthermore, the scale
 $\Lambda_R$ is constrained to be at least $10$ TeV, in order to comply with experimental bound arising from the singly-charged extra gauge boson. The additional VEVs, $w$ and $w_R$, are taken to be of order several TeV. Other VEVs are neglected in this analysis, as they do not play a role in the subsequent discussion.\\
 
We begin our numerical investigation by revisiting the phenomenology of the light neutrino contribution to neutrinoless double beta ($0\nu\beta\beta$) decay, which has been extensively studied before \cite{Tanabashi:2018oca}. In the case of a normal ordering (NO) neutrino mass spectrum, the magnitude of the effective Majorana mass, denoted as $|m_{eff,\nu_L}^{LL}|$, exhibits a strong dependence on the lightest neutrino mass. For $m_1\leq 8\times 10^{-4}$ eV, corresponding to $m_1\ll m_2<m_3$, the value of $|m_{eff,\nu_L}^{LL}|$ lies within the interval $(0.4-4.8)\times 10^{-3}$ eV. When 
$m_1$ is in the range $(10^{-4}-10^{-3})$ eV, the effective mass can be significantly suppressed depending on the mixing angles within their allowed $3\sigma$ ranges and the CP-violating phases (both Dirac and Majorana). For $m_1\geq 10^{-2}$ eV, 
$|m_{eff,\nu_L}^{LL}|$ increases approximately linearly with 
$m_1$, and in the quasi-degenerate regime ($m_1\geq 0.1$ eV), one obtains $|m_{eff,\nu_L}^{LL}|\geq0.05$ eV.\\

\begin{figure}
\begin{center}
\begin{tabular}{c}
\includegraphics[width=10cm,height=6.5cm]{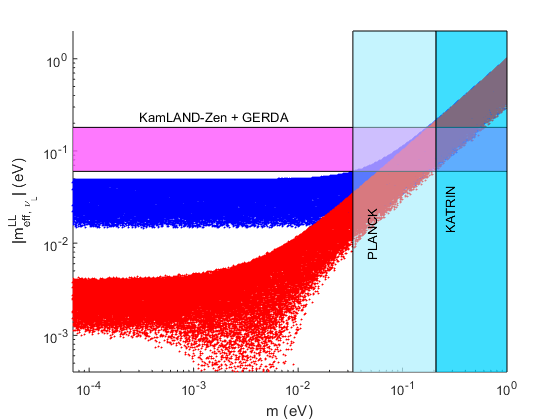}
\end{tabular}
\caption{The $0\nu\beta\beta$ effective mass parameter, arising from the fundamental channel as a function of the lightest neutrino mass 
$m$, is evaluated for both normal ordering (NO) and inverted ordering (IO) neutrino mass spectra, represented by red 
and blue scatter points, respectively.}
\label{FundaContri}
\end{center}
\end{figure}

In contrast to the previously discussed scenario, the behavior of the effective Majorana mass parameter $|m_{eff,\nu_L}^{LL}|$ under the inverted ordering (IO) neutrino mass spectrum, characterized by $m_3<m_1<m_2$, exhibits a more monotonic trend. For the regime $m_3^2\ll \Delta m_{23}^2$, corresponding to $m_3<1.6\times 10^{-2}$ eV, the value of $|m_{eff,\nu_L}^{LL}|$ becomes independent of $m_3$. In this domain, $|m_{eff,\nu_L}^{LL}|$ remains confined within the interval
$(1.5-5.0)\times 10^{-2}$ eV. As $m_3$ increases beyond $1.6\times 10^{-2}$ eV, the magnitude of 
$|m_{eff,\nu_L}^{LL}|$ exhibits a mild growth. Upon entering the quasi-degenerate regime, 
defined roughly by $m_3\geq 0.1$ eV, $|m_{eff,\nu_L}^{LL}|$ increases approximately linearly with
$m_3$. The detailed dependence of $|m_{eff,\nu_L}^{LL}|$ on the lightest neutrino mass is illustrated in Fig. \ref{FundaContri}.\\

To estimate the relative contributions of the channels defined in Eq. (\ref{etadef}), it is convenient to recast the parameter 
$\eta$ in terms of an effective Majorana mass. We define:
\be
m_{eff,\alpha}^{\beta}=\left|\eta_\alpha^{\beta}\right|m_e\left|\frac{{\cal M}^{0\nu}_{\alpha}}{{\cal M}^{0\nu}_{\nu_L}}\right|,
\ee
where the index $\alpha$ denotes $\nu_{L(R)}$, $H_i$ ($i=1,2$), and 
$\beta=LL, RR, LR, RL$, respectively. The magnitudes of the couplings 
$V_L^{W_{1\mu}}$ and $V_R^{W_{2\mu}}$, which play a crucial role in determining the decay rate, can be approximated as follows:
\be
\left(V_L^{W_{1\mu}}\right)_{ei}\sim (U_A)_{ei}\sim \left(m_DM_R^{-1}U_\nu^{M}\right)_{1i}
\sim\frac{u}{u'}\left(m_l\right)_{1k}\left(M_R^{-1}\right)_{ki}\sim\frac{u}{u'}\frac{m_e}{M_R}<2.0\times 10^{-10}.
\label{Vlimit01}
\ee
\be
\left(V_R^{W_{2\mu}}\right)_{ei}\sim (U_B)_{ei}\sim\left(M_R^{-1}m_D^TU_\nu\right)_{1i}
\sim\frac{u}{u'}\left(M_R^{-1}\right)_{1k}\left(m_l\right)_{ki}\sim\frac{u}{u'}\frac{m_\tau}{M_R}<7.2\times 10^{-7}.
\label{Vlimit02}
\ee
In deriving Eqs. (\ref{Vlimit01}) and (\ref{Vlimit02}), we have used the fact that the charged lepton mass matrix $m_l$
is diagonal in the chosen basis, with the tau lepton mass dominating over those of the electron and muon, 
as precisely determined by the experiments. To obtain the results, we have set $u=0.1$ GeV, $u'=246$ GeV and $M_R=1$ TeV.\\

We first examine the contribution arising from the exchange channel involving the singly charged gauge boson $W_1$
and the mediation of heavy right-handed neutrinos, denoted by $\eta_{\nu_R}^{LL}$. The upper limit for effective mass parameter 
in this case is given by
\be
m_{eff,\nu_R}^{LL}=\frac{\left(V_L^{W_{1\mu}}\right)_{ei}^2 m_p m_e}{M_i}\left|\frac{{\cal M}^{0\nu}_{\nu_R}}{{\cal M}^{0\nu}_{\nu_L}}\right|<4.7\times 10^{-12} {\rm eV}.
\ee
Here, the upper bound on $V_L^{W_{1\mu}}$ is determined from Eq. (\ref{Vlimit01}), while the heavy neutrino masses are set at 
$M_i=1$ TeV, ($i=1,2,3$). The ratio of nuclear matrix elements ${\cal M}^{0\nu}_{\nu_R}/{\cal M}^{0\nu}_{\nu_L}$
 is taken at its maximum for the isotope $^{76}{\rm Ge}$, with values in the range 
 $22.2\leq{\cal M}^{0\nu}_{\nu_R}/{\cal M}^{0\nu}_{\nu_L}\leq 76.3$, as indicated in Table \ref{tab:nucl-matrix}.\\
 
 Applying the same methodology to other channels, and recalling that the light neutrino mass satisfies 
 $m_i<1$ eV, with $\Lambda_L< 1$ eV constrained by neutrino mass bounds, and setting the physical scalar Higgs mass at 
$M_{H_1}\simeq 1$ TeV, we obtain the following effective mass contributions:
\be
m_{eff,\nu_L}^{RR}=\left(\frac{M_{W_1}}{M_{W_2}}\right)^4\left(V_R^{W_{2\mu}}\right)_{ei}^2 m_i<3.4\times 10^{-20} {\rm eV},
\ee
\be
m_{eff,\nu_L}^{LR}=\left(\frac{M_{W_1}}{M_{W_2}}\right)^2\left(U_L^{W_{1\mu}}\right)_{ei}
\left(V_R^{W_{2\mu}}\right)_{ei} m_i<1.8\times 10^{-10} {\rm eV},
\ee
\be
m_{eff,\nu_L}^{RL}=\left(\frac{M_{W_1}}{M_{W_2}}\right)^2\frac{
\left(V_L^{W_{1\mu}}\right)_{ei}\left(U_R^{W_{2\mu}}\right)_{ei}m_p m_e}{M_i} 
\left|\frac{{\cal M}^{0\nu}_{\nu_R}}{{\cal M}^{0\nu}_{\nu_L}}\right|<1.9\times 10^{-9} {\rm eV},
\ee
\be
m_{eff,H_1}=\frac{\Lambda_L m_p m_e}{M_{H_1}^2}<5.0\times 10^{-10} {\rm eV}.
\ee
All of these contributions are found to be exceedingly small compared to the dominant light neutrino exchange mechanism, 
both for the normal ordering (NO) and inverted ordering (IO) scenarios. Furthermore, they lie far below the current experimental 
sensitivities (see Table \ref{tab:halflife}). Consequently, while these channels provide negligible corrections, other contributions-specifically those associated with $\eta_{\nu_R}^{RR}$ and $\eta_{H_2}$-are expected to play a more significant 
role and will be discussed in detail in the later parts.\\

\begin{figure}
\begin{center}
\begin{tabular}{cc}
\includegraphics[width=8cm,height=5.5cm]{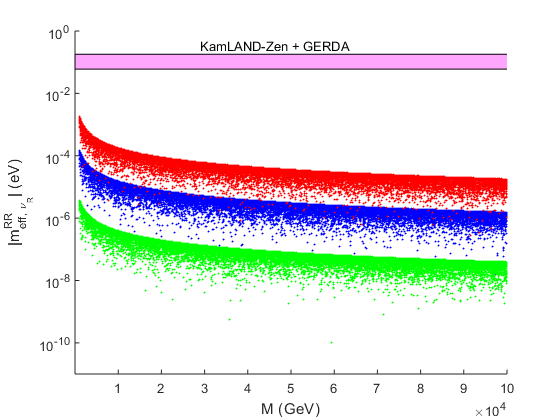}&
\includegraphics[width=8cm,height=5.5cm]{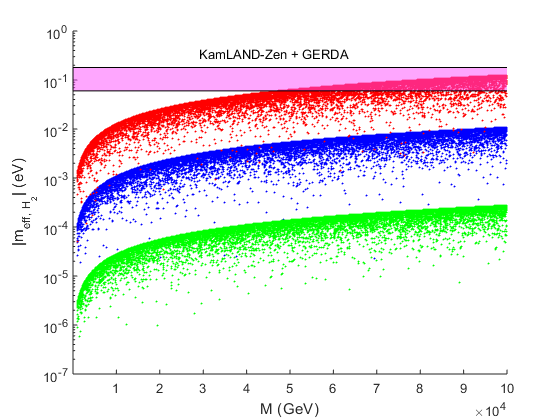}
\end{tabular}
\caption{The $0\nu\beta\beta$ effective mass parameter as function of the heavy neutrino mass $M$, contributed by heavy gauge 
boson and neutrino exchanges (left panel); and by the mediation of doubly charged Higgs scalar $H_2^{--}$, for three different values
of $\Lambda_R=10.8,20,50$ TeV (represented by red, blue, and green points, respectively).}
\label{FigRRH2}
\end{center}
\end{figure}

The effective mass parameter for the channel involving heavy gauge boson and neutrino exchanges,
$|m_{eff,\nu_R}^{RR}|\sim |\eta_{\nu_R}^{RR}|\sim 1/(M_{W_2}^4M_i)\sim 1/(\Lambda_{R}^4M_i)$,
is inversely proportional to $M_i$ and $\Lambda_R^4$, and therefore decreases rapidly with increasing 
$\Lambda_R$. The dependence of $|m_{eff,\nu_R}^{RR}|$ on $M$, representing the masses of the three heavy neutrinos 
(assumed equal for simplicity, $M_i=M$, $i=1,2,3$), is illustrated in the left panel of Figure \ref{FigRRH2}. 
The figure is plotted for three different values of 
$\Lambda_R=10.8,~20,~50$ TeV, corresponding to the red, blue, and green points, respectively, with 
$M$ varying in range of $(1-100)$ TeV. From the figure, it is evident that at 
$M=1$ TeV, $\Lambda_R=10.8$ TeV, the maximum value of the effective mass parameter can reach 
$|m_{eff,\nu_R}^{RR}|=1.6\times 10^{-3}$ eV. This value is sizable compared to the contribution of the 
fundamental channel in the case of the normal ordering (NO) neutrino mass spectrum, particularly in the region 
below the quasi-degenerate regime ($m_1\leq 0.1$ eV).\\

The dependence of the effective mass parameter on the heavy neutrino mass for the channel mediated by the physical Higgs scalar 
$H_2^{--}$ is illustrated in the right panel of Figure \ref{FigRRH2}. The relation 
$|m_{eff,H_2}|\sim |\eta_{H_2}|\sim (Y_{H_2}^R\Lambda_R)/(M_{W_R}^4M_{H_2}^2)\sim M/(\Lambda_R^4M_{H_2}^2)$
demonstrates that the parameter is directly proportional to the heavy neutrino mass 
$M$. Consequently, for fixed values of $\Lambda_R=10.8,20,50$ TeV (represented by red, blue, and green points, respectively), 
$|m_{eff,H_2}|$ increases linearly with $M$. The magnitude of $|m_{eff,H_2}|$ can significantly impact both 
the normal ordering (NO) and inverted ordering (IO) neutrino mass spectra, particularly when $\Lambda_R$ 
 lies in the lower interval of $(10-15)$ TeV. For example, at $\Lambda_R=10.8$ TeV, the value of 
$|m_{eff,H_2}|$ exceeds the experimental limits set by KamLAND-Zen and GERDA for $M>45$ TeV. At the same 
 $\Lambda_R$, our calculations indicate that the maximum values of $|m_{eff,H_2}|$ can reach 
$10^{-2};~5\times 10^{-3}$ eV for $M=7.2;~ 4.5$ TeV, respectively. These values are approximately five times larger than 
those obtained for $|m_{eff,\nu_R}^{RR}|$. All calculations are performed under the assumption $M_{H_2^{--}}=M_{H}=1$ TeV.\\

\begin{figure}
\begin{center}
\begin{tabular}{cc}
\includegraphics[width=8cm,height=5.5cm]{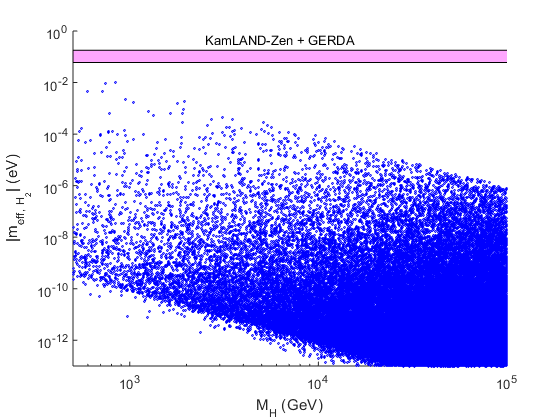}&
\includegraphics[width=8cm,height=5.5cm]{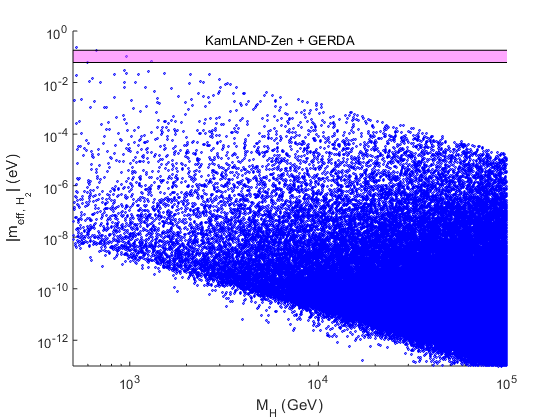}
\end{tabular}
\caption{
The dependence of the effective Majorana mass $|m_{eff,H_2}|$, on the doubly charged Higgs boson mass 
$M_{H_2^--}=M_H$ for two representative values of the heavy mass scale: 
$M=5\times 10^{3}$ GeV (left panel) and $M=5\times 10^{4}$ GeV (right panel).}
\label{FigH2}
\end{center}
\end{figure}
We examine the dependence of $|m_{eff,H_2}|$ on the Higgs mass, as shown in Figure \ref{FigH2}, for two fixed 
values of the heavy neutrino mass: $M=5\times 10^{3}$ GeV (left panel) and $M=5\times 10^{4}$ GeV (right panel).
 The Higgs mass, $M_{H_2}=M_H$, is varied from 500 GeV up to 100 TeV, while the new physics scale, $\Lambda_R$, 
 is randomly scanned within the interval $(10-1000)$ TeV. The results indicate that the magnitude of 
$|m_{eff,H_2}|$ decreases rapidly with increasing values of both 
$M$ and $\Lambda_R$. However, in the lower ranges of these parameters, the Higgs-mediated channel can yield 
contributions to the effective mass parameter comparable to those of the fundamental light-neutrino mechanism.\\
\begin{figure}
\begin{center}
\begin{tabular}{c}
\includegraphics[width=10cm,height=6.5cm]{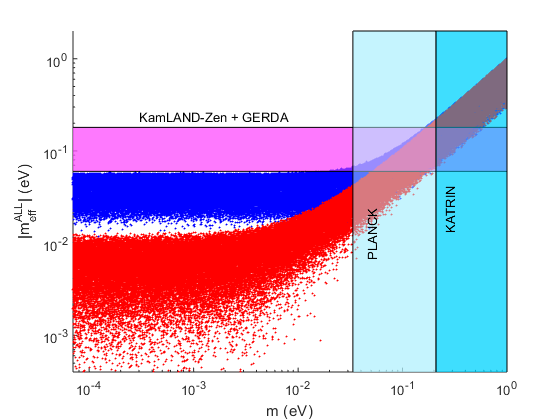}
\end{tabular}
\caption{The total effective mass parameter $|m_{eff}^{ALL}|$, contributed by all leading channels as a function 
of the lightest neutrino mass $m$, for both normal ordering (NO) and inverted ordering (IO) neutrino mass spectra, corresponding to red 
and blue scatter points, respectively.}
\label{TotalContri}
\end{center}
\end{figure}

The total Majorana effective mass parameter, $|m_{eff}^{ALL}|$, which is provided by 
all of the channels described in the Figures \ref{GaugeExc} and \ref{HiggExc}, is visually 
shown in the Figure \ref{TotalContri} as function of the lightest neutrino mass. The contributions are, in fact, absolutely dominated by the fundamental channel and the channel with mediation of doubly-charged Higgs scalar $H_2^{--}$. To make the plots, we have used 
$\Lambda_R=10.8$ TeV, $M_H=1$ TeV and $M=5$ TeV for both normal ordering (NO) and 
inverted ordering (IO)  neutrino mass spectrum.\\

In the case of the normal neutrino mass spectrum (represented by red scatter points), the most evident feature is the alteration in the overall pattern shape. Unlike the fundamental scenario, where suppression of the effective mass occurs only within the narrow interval 
$m\in(10^{-4}-10^{-3})$ eV, the quantity $|m_{eff}^{ALL}|$ experiences significant suppression for values of 
$m\leq 3\times 10^{-3}$ eV. For $m\leq 3\times 10^{-3}$ eV, the maximum value of $|m_{eff}^{ALL}|$ corresponding to a fixed 
$m$ remains nearly constant, with a magnitude of approximately $1.4\times 10^{-2}$ eV, that is about three times larger than those 
of $|m_{eff,\nu_L}^{LL}|$ of the channel with light neutrino mediation only. Beyond this threshold, the 
effective mass exhibits a gradual increase, eventually entering the quasi-degenerate regime, where it rises linearly 
in a manner analogous to the fundamental case.\\

In the inverted order neutrino mass spectrum, the pattern shape  of the total effective Majorana mass $|m_{eff}^{ALL}|$ 
(blue scatter points) is rather similar to the $|m_{eff,\nu_L}^{LL}|$ of the fundamental case. In the inverted hierarchy 
interval of the light neutrino mass $m_3^2\ll \Delta m_{23}^2$, corresponding to $m_3\leq 7\times 10^{-3}$ eV, the value of $|m_{eff}^{ALL}|$ is virtually independent of $m_3$, which lies in the range of $(1.5-5.7)\times 10^{-2}$ eV. The maximum of this band is close to the  
KamLAND-Zen and GERDA limits and little higher than those of $|m_{eff,\nu_L}^{LL}|$ in fundamental case, which gets 
value in the interval $(1.5-5.0)\times 10^{-2}$ eV. Upon entering the quasi-degenerate area with $m3\geq 0.1$ eV, the both cases show the same behavior with linearly increasing as function of $m_3$.

\section{\label{conl} Conclusion}
In this research, we have briefly reviewed the flipped trinification model with gauge symmetry 
$SU(3)_C\otimes SU(3)_L\otimes SU(3)_R \otimes U(1)_X$ and performed a numerical analysis of the generalized effective Majorana mass in neutrinoless double beta $(0\nu\beta\beta)$
 decay. The results indicate that, among the channels involving new physics, only the exchange of heavy singly-charged gauge bosons accompanied by new neutrino mediation, and the mediation of the doubly-charged Higgs scalar $H_2^{--}$, provide meaningful contributions to the total effective Majorana mass.\\
 
The contribution from the channel with heavy gauge boson and massive neutrino exchanges, characterized by $|m_{eff,\nu_R}^{RR}|$, decreases rapidly with increasing new physics scale 
$\Lambda_R$ and heavy neutrino mass $M$. However, when $\Lambda_R$ and $M$ are not too large, 
the contribution remains sizable. For instance, at $M=1$ TeV and $\Lambda_R=10.8$ TeV, 
the maximum of $|m_{eff,\nu_R}^{RR}|$ can reach $1.6\times 10^{-3}$ eV, which is non-negligible, especially in the case of normal ordering (NO) neutrino mass spectrum.\\

The most significant new physics contribution arises from the channel mediated by the doubly-charged scalar $H_2^{--}$. The corresponding quantity $|m_{eff,H_2}|$ is inversely proportional to 
$\Lambda_R^4$, but at fixed $\Lambda_R$ it increases linearly with the heavy neutrino mass 
$M$. This contribution can be comparable to the fundamental mechanism. For 
$\Lambda_R=10.8$ TeV, the maximum values of $|m_{eff,H_2}|$ may reach 
$5\times 10^{-3}$ and $10^{-2}$ eV at $M=4.5$ and $7.2$ TeV, respectively. In fact, the effective mass can exceed the limits set by KamLAND-Zen and GERDA, offering the possibility of constraining heavy neutrino masses.\\

As illustrated in Figure \ref{TotalContri}, the total effective Majorana mass pattern, incorporating all significant contributions, exhibits considerable deviation from the fundamental case. For the normal ordering spectrum, suppression of $|m_{eff}^{ALL}|$ occurs for $m_1\leq 3\times 10^{-3}$
 eV, instead of being confined to the narrower interval $(10^{-4}-10^{-3})$
, and its maximum value is approximately three times higher, reaching 
$1.4\times 10^{-2}$ eV. In the inverted ordering scenario, the overall pattern remains similar, though the maximum value increases slightly to $5.7\times 10^{-2}$ eV compared to 
$5.0\times 10^{-2}$ eV in the fundamental case.

\section*{Acknowledgments}
This research is funded by the Vietnam Academy of Science and Technology (VAST) under
Grant No. CSCL05.01/25-26.
	
\bibliography{combine}

\end{document}